\documentclass[12pt]{article}
\usepackage{amsmath}
\usepackage{graphics,graphicx,xcolor}

\def\be{\begin{equation}}
\def\ee{\end{equation}}
\def\arr{\begin{array}{rll}}
\def\ea{\end{array}}
\def\bea{\begin{eqnarray}}
\def\eea{\end{eqnarray}}

\def\N2{$N{=}2$}

\def\>{\rangle}
\def\<{\langle}
\def\+{\dagger}
\def\={\ =\ }

\def\bal{\begin{aligned}}
\def\eal{\end{aligned}}
\begin{document}
\begin{titlepage}
\setcounter{page}{0}
\begin{center}
{\LARGE\bf  Equations of fluid dynamics with the }\\
\vskip 0.5cm
{\LARGE\bf $\ell$--conformal Galilei symmetry}\\
\vskip 1.5cm
\textrm{\Large Anton Galajinsky \ }
\vskip 0.7cm
{\it
Tomsk State University of Control Systems and Radioelectronics, 634050 Tomsk, Russia} \\

\vskip 0.2cm
{e-mail: a.galajinsky@tusur.ru}
\vskip 0.5cm
\end{center}

\begin{abstract} \noindent
Equations of fluid dynamics are formulated, which hold invariant under the action of the $\ell$--conformal Galilei group. They include the conventional continuity equation, a higher order material derivative analogue of the Euler equation, and a suitable modification of the conventional equation of state. Conserved charges associated with the $\ell$--conformal Galilei symmetry transformations are presented.
\end{abstract}

\vspace{0.5cm}

\end{titlepage}
\renewcommand{\thefootnote}{\arabic{footnote}}
\setcounter{footnote}0

\noindent
{\bf 1. Introduction}\\

\noindent
Recent exploration of the nonrelativistic variant of the AdS/CFT--correspondence extended the holographic dictionary to strongly coupled condensed matter systems (for a review see \cite{NS} and references therein). From a symmetry standpoint, the central object here is a finite--dimensional conformal extension of the Galilei algebra \cite{Henkel,NOR}.

Transformations forming the corresponding group include
(temporal) translation, dilatation, and special
conformal transformation, which form $SL(2,R)$ subgroup, as well as spatial rotations, spatial translations, Galilei boosts and constant accelerations. Structure relations of the associate Lie algebra involve an arbitrary (half)integer parameter $\ell$, which specifies the number of acceleration generators at hand \cite{Henkel,NOR}. For this reason, the algebra is commonly referred to as the $\ell$--conformal Galilei algebra.

A peculiar feature of the nonrelativistic conformal transformations is that temporal and spatial coordinates scale differently under the dilatation: $t'=\lambda t$, $x'_i=\lambda^{\ell} x_i$. In particular, the reciprocal of $\ell$ is known as the rational dynamical exponent and it is also customary to refer to the instances of $\ell=\frac 12$ and $\ell=1$ as the Schr\"odinger group and the conformal Galilei group, respectively.

Extensive recent studies of dynamical realisations of the $\ell$--conformal Galilei group \cite{LSZ1}--\cite{ME} revealed interesting peculiarities. Because a number of functionally independent integrals of motion needed to integrate a differential equation correlates with its order, such systems in general involve higher derivative terms \cite{LSZ1,DH1,GK,AGM,AGKM,AGGM,GM4,M,KLS,IM1}. In particular, the Pais--Uhlenbeck oscillator \cite{PU} enjoys such a symmetry for a special choice of its frequencies \cite{AGGM,KLS,IM1}. Second order systems, for which conserved charges associated with the acceleration generators
are functionally dependent, have been constructed in \cite{GM3,GM1} (see also \cite{FIL}). Models without higher derivative terms, for which all constants of motion are functionally independent, were built in \cite{CG,DC} by making recourse to geodesics on Ricci–-flat or Einstein spacetimes with the $\ell$--conformal Galilei isometry
group. A peculiar feature of those systems is that the dimension of spacetime grows with the value of $\ell$.

As is well known, the Euler equations describing a perfect fluid enjoy the $\ell=\frac 12$ conformal Galilei symmetry provided a specific equation of state, which links pressure to density, is imposed \cite{RS,JNPP} . Realisations of the $\ell=\frac 12$ and $\ell=1$ conformal Galilei transformations within the context of fluid mechanics have been extensively studied in the past (see e.g. \cite{RS}--\cite{HZ} and references therein\footnote{Note that some statements in the literature regarding conformal symmetries of fluid mechanics contradict each other. For a detailed clarification see \cite{HZ}.}).
It is then natural to wonder whether the Euler equations and the equation of state can be modified so as to accommodate the $\ell$--conformal Galilei symmetry for an arbitrary value of the parameter $\ell$. The goal of this work is to construct such a generalisation.

The paper is organised as follows. In the next section, an action of the $\ell$--conformal Galilei group upon a density scalar field and a velocity vector field is established. In Sect. 3, equations of motion are formulated, which include the conventional continuity equation and a higher order material derivative analogue of the Euler equation. A modification of the conventional equation of state is found, which ensures the $\ell$--conformal Galilei symmetry of the resulting system of partial differential equations. Conserved charges associated with the $\ell$--conformal Galilei symmetry transformations are discussed in Sect. 4. While the acceleration charges can be constructed both for integer and half--integer $\ell$, the $SL(2,R)$--charges and the energy--momentum tensor are readily obtained for a half--integer $\ell$ only. In the concluding Sect. 5 we summarise our results and discuss possible further developments. Infinitesimal form of the $\ell$--conformal Galilei transformations acting upon spacetime coordinates, the density and the velocity vector field is given in Appendix.

Throughout the paper, summation over repeated indices is understood unless otherwise is stated.

\vspace{0.5cm}

\noindent
{\bf 2. The $\ell$--conformal Galilei group}\\

\noindent
Consider a nonrelativistic spacetime parametrised by a temporal variable $t$ and Cartesian coordinates $x_i$, $i=1,\dots,d$. The $\ell$--conformal Galilei group involves the $SL(2,R)$--transformations
\be\label{sl2}
t'=\frac{\alpha t+\beta}{\gamma t+\delta}, \qquad
x'_i={\left(\frac{\partial t'}{\partial t} \right)}^\ell x_i,
\ee
where $\alpha \delta-\beta \gamma=1$ and $\ell$ is an arbitrary (half)integer number, and a chain of transformations with the vector parameters $a^{(0)}_i$, $a^{(1)}_i$, $\dots$, $a^{(2\ell)}_i$ (no sum over repeated index $n$)
\be\label{acs}
t'=t, \qquad x'_i=x_i+a^{(n)}_i t^n,
\ee
where $n=0,\dots, 2\ell$. In the preceding formula, $n=0$ and $n=1$ correspond to the spatial translation and the Galilei boost, while higher values of $n$ describe $n$--th order constant accelerations. The group also contains conventional $SO(d)$--rotation, which in what follows will be disregarded. The elementary relations
\be\label{suppl}
t=\frac{\beta-\delta t'}{\gamma t'-\alpha}, \qquad  \frac{\partial t}{\partial t'}=\frac{1}{{(\gamma t'-\alpha)}^2}, \qquad \frac{\partial t'}{\partial t}=\frac{1}{{(\gamma t+\delta)}^2}={(\gamma t'-\alpha)}^2,
\ee
will prove useful below.

The generators of infinitesimal (temporal) translation, dilatation, special conformal transformation, and accelerations follow from (\ref{sl2}) and (\ref{acs}) (for more details see Appendix)
\be
H=\frac{\partial}{\partial t}, \qquad D=t \frac{\partial}{\partial t}+\ell x_i \frac{\partial}{\partial x_i}, \qquad K=t^2 \frac{\partial}{\partial t}+2 \ell t x_i \frac{\partial}{\partial x_i}, \qquad C^{(n)}_i=t^n  \frac{\partial}{\partial x_i}.
\ee
They obey structure relations the $\ell$--conformal Galilei algebra \cite{Henkel,NOR}
\begin{align}\label{algebra}
&
[H,D]=H, &&  [H,K]=2 D, && [D,K]=K,
\nonumber\\[2pt]
&
[H,C^{(n)}_i]=n C^{(n-1)}_i, && [D,C^{(n)}_i]=(n-l) C^{(n)}_i, && [K,C^{(n)}_i]=(n-2l) C^{(n+1)}_i,
\end{align}
where $i=1,\dots,d$, $n=0,\dots, 2\ell$, and $\ell$ is an arbitrary (half)integer real parameter. Note that it is the latter restriction which guarantees that the algebra is finite--dimensional.

Let $\rho(t,x)$ and $\upsilon_i (t,x)$, $i=1,\dots,d$, be the density and the velocity vector field which characterise a fluid. Transformation law of $\rho(t,x)$ under the action of the $\ell$--conformal Galilei group is obtained by fixing a value of the temporal variable $t$ and demanding the mass of a $d$--dimensional volume element $V$ to be invariant under (\ref{sl2}), (\ref{acs})
\be
\int_{V'} d x' \rho' (t',x')=\int_{V} d x \rho(t,x),
\ee
where $dx=dx_1 \dots dx_d$.
This gives
\be\label{trr}
\rho(t,x)= {\left(\frac{\partial t'}{\partial t} \right)}^{\ell d} \rho' (t',x'),
\ee
for the $SL(2,R)$--transformation and
\be\label{trr1}
\rho(t,x)=\rho' (t',x'),
\ee
for the accelerations.

Similarly,
considering an orbit of a liquid particle parametrised by $x_i (t)$ and taking into account the relation
\be
\frac{d x_i (t)}{d t}= \upsilon_i (t,x(t)),
\ee
from eqs. (\ref{sl2}), (\ref{acs}) one gets\footnote{As is usual in classical dynamics, when restricting the transformations similar to (\ref{sl2}), (\ref{acs}) to a particle orbit, one replaces $x'_i$ with $x'_i (t')$ and $x_i$ with $x_i (t)$.} (no sum over repeated index $n$)
\be\label{trv}
\upsilon_i (t,x)={\left(\frac{\partial t'}{\partial t} \right)}^{1-\ell} \upsilon'_i (t',x')+\frac{\partial}{\partial t} {\left(\frac{\partial t'}{\partial t} \right)}^{-\ell} x'_i,
\ee
for the $SL(2,R)$--transformation and
\be\label{trv1}
\upsilon_i (t,x)=\upsilon'_i (t',x')-n a^{(n)}_i t^{n-1},
\ee
for the accelerations.

For the readers's convenience, infinitesimal form the $\ell$--conformal Galilei transformations acting upon the coordinates and the fields is displayed in Appendix.

\vspace{0.5cm}

\noindent
{\bf 3. Equations of fluid dynamics with the $\ell$--conformal Galilei symmetry}\\

\noindent
Having fixed transformation laws of the density and the velocity vector field, we are now in a position to formulate equations of fluid dynamics with the $\ell$--conformal Galilei symmetry.

Because the continuity equation has a clear physical meaning (it describes the mass transport), it should be kept intact
\be\label{ce}
\frac{\partial \rho}{\partial t} + \frac{\partial ( \rho \upsilon_i )}{\partial x_i}=0.
\ee
Taking into account the identities
\be
\frac{\partial}{\partial t}=\left(\frac{\partial t'}{\partial t}\right) \frac{\partial}{\partial t'}+\left(\frac{\partial x'_i}{\partial t} \right) \frac{\partial}{\partial x'_i},
\qquad \frac{\partial}{\partial x_i}=\left( \frac{\partial t'}{\partial x_i} \right) \frac{\partial}{\partial t'}+\left(\frac{\partial x'_j}{\partial x_i} \right)\frac{\partial}{\partial x'_j},
\ee
and eqs. (\ref{sl2}), (\ref{acs}), (\ref{trr}), (\ref{trr1}), one readily gets
\be
\frac{\partial \rho}{\partial t} + \frac{\partial ( \rho \upsilon_i )}{\partial x_i}= \frac{\partial \rho'}{\partial t'} + \frac{\partial ( \rho' \upsilon'_i )}{\partial x'_i}
\ee
for the acceleration transformations, and
\be
\frac{\partial \rho}{\partial t} + \frac{\partial ( \rho \upsilon_i )}{\partial x_i}= {\left(\frac{\partial t'}{\partial t}\right)}^{\ell d+1} \left( \frac{\partial \rho'}{\partial t'} + \frac{\partial ( \rho' \upsilon'_i )}{\partial x'_i}\right)
\ee
for the conformal transformations.
Thus, the transport equation (\ref{ce}) holds invariant under the action of the $\ell$--conformal Galilei group.

In order to formulate an analogue of the Euler equation, one introduces the material derivative
\be
\mathcal{D}=\frac{\partial}{\partial t} +\upsilon_i  (t,x) \frac{\partial}{\partial x_i},
\ee
and studies transformation laws of $\mathcal{D} \upsilon_i$, $\mathcal{D}^2 \upsilon_i$ etc. under (\ref{sl2}) and (\ref{acs}).
Let us start with $\mathcal{D} \upsilon_i$. Given the conformal transformations
(\ref{sl2}), the material derivative transforms covariantly
\be
\mathcal{D}=\left(\frac{\partial t'}{\partial t} \right) \mathcal{D}'.
\ee
Taking into account eq. (\ref{trv}), one obtains
\be\label{DV}
\mathcal{D} \upsilon_i={\left(\frac{\partial t'}{\partial t} \right)}^{2-\ell} \mathcal{D}' \upsilon'_i+(1-2\ell) {\left(\frac{\partial t'}{\partial t} \right)}^{-\ell} \left(\frac{\partial^2 t'}{ \partial t^2} \right)  \upsilon'_i  +\frac{\partial^2}{ \partial t^2} {\left(\frac{\partial t'}{\partial t} \right)}^{-\ell}  x'_i,
\ee
where $\upsilon_i=\upsilon_i (t,x)$, $\upsilon'_i=\upsilon'_i (t',x')$. Although for arbitrary value of $\ell$ the field $\mathcal{D} \upsilon_i$ does not transform covariantly, the second term drops out for $\ell=\frac 12$, while the third term
\be
\frac{\partial^2}{ \partial t^2} {\left(\frac{\partial t'}{\partial t} \right)}^{-\ell}=-\ell  {\left(\frac{\partial t'}{\partial t} \right)}^{-\ell} \left(\frac{\dddot{t'}}{\dot t'}-(\ell+1){\left(\frac{\ddot t'}{\dot t'} \right)}^2 \right),
\ee
where the dot designates the derivative with respect to $t$, is proportional to the Schwarzian derivative
\be
S[t'(t)]=\frac{\dddot{t'}}{\dot t'}-\frac 32 {\left(\frac{\ddot t'}{\dot t'} \right)}^2,
\ee
whenever $\ell=\frac 12$.
As is well known, the latter vanishes identically for $t'$ in (\ref{sl2}).

Thus, for $\ell=\frac 12$ the material derivative $\mathcal{D} \upsilon_i$ transforms covariantly under the $SL(2,R)$--transformation.
In a similar fashion, for an arbitrary (half)integer value of $\ell$ one can establish the relation\footnote{In obtaining eq. (\ref{Rel}), it proves useful to focus on a concrete orbit $x_i (t)$ such that $\mathcal{D}=\left(\frac{\partial t'}{\partial t} \right) \mathcal{D}'={(\gamma t'-\alpha)}^2 \frac{d}{d t'}$, $\upsilon_i (t, x(t))={(\gamma t'-\alpha)}^{2(1-\ell)} \frac{d x'_i (t')}{d t'} -2 \ell \gamma {(\gamma t'-\alpha)}^{1-2 \ell} x'_i (t')$, and then compute $\mathcal{D}^n \upsilon_i$ for an arbitrary integer $n$.}
\be\label{Rel}
\mathcal{D}^{2\ell} \upsilon_i={\left(\frac{\partial t'}{\partial t} \right)}^{\ell+1} {\mathcal{D}'}^{2\ell} \upsilon'_i.
\ee

Turning to the acceleration transformations (\ref{acs}), (\ref{trv1}), one readily gets\footnote{A straightforward calculation gives (no sum over repeated index $n$) ${\mathcal{D}}^{2\ell}  \upsilon_i={\mathcal{D}'}^{2\ell} \upsilon'_i-(2\ell+1)! C_n^{2\ell+1} a^{(n)}_i t'^{n-2\ell-1}$, $C_n^{2\ell+1}$ being the binomial coefficients and $n=0,\dots,2\ell$. Given a (half)integer $\ell$ and $n=0,\dots,2\ell$, all $C_n^{2\ell+1}$ prove to vanish.}
\be
\mathcal{D}=\mathcal{D}', \qquad {\mathcal{D}}^{2\ell} \upsilon_i={\mathcal{D}'}^{2\ell} \upsilon'_i,
\ee
where $\upsilon_i=\upsilon_i (t,x)$, $\upsilon'_i=\upsilon'_i (t',x')$.

As a final step, let us introduce the pressure $p(t,x)$, which is subject to the equation of state  \cite{RS,JNPP}
\be\label{eos}
p=\nu \rho^\mu,
\ee
where $\nu$ and $\mu$ are real constants, and then impose the higher derivative analogue of the Euler equation
\be\label{ee}
\rho  \mathcal{D}^{2\ell} \upsilon_i=-\frac{\partial p}{\partial x_i}.
\ee
The latter automatically holds invariant under the acceleration transformations, while for the conformal transformations one finds
\be
{\left(\frac{\partial t'}{\partial t} \right)}^{\ell d+1 } \rho'  \mathcal{D}'^{2\ell} \upsilon'_i=-{\left(\frac{\partial t'}{\partial t} \right)}^{\ell d \mu }  \frac{\partial p'}{\partial x'_i}.
\ee
Imposing the restriction on the exponent
\be
\mu=1+\frac{1}{\ell d},
\ee
one ensures the $SL(2,R)$--invariance.

To summarise, generalised perfect fluid equations, which hold invariant under the action of the $\ell$--conformal Galilei group, read
\be\label{fin}
\frac{\partial \rho}{\partial t} + \frac{\partial ( \rho \upsilon_i )}{\partial x_i}=0, \qquad \rho  \mathcal{D}^{2\ell} \upsilon_i=-\frac{\partial p}{\partial x_i}, \qquad p=\nu \rho^{1+\frac{1}{\ell d}},
\ee
where $\ell$ is a (half)integer number, $d$ is the dimension of space, $\mathcal{D}=\frac{\partial}{\partial t} +\upsilon_i \frac{\partial}{\partial x_i}$, and
$\nu$ is a constant. It should be mentioned that the $\ell=\frac 12$ case was studied in detail in \cite{RS,JNPP}.

In the next section, we shall construct conserved charges associated with the $\ell$--conformal Galilei symmetry of the equations of motion (\ref{fin}). To that end, it proves useful to introduce the potential function $V(\rho)$, which via the Legendre transform gives the pressure \cite{JNPP}
\be\label{pot}
p(\rho)=\rho V'(\rho)-V(\rho).
\ee
Given the equation of state $p=\nu \rho^{1+\frac{1}{\ell d}}$, the potential reads
\be\label{pot1}
V=\ell d p+w \rho,
\ee
where $d$ is the dimension of space and $w$ is a constant of integration. As shown below, $V$ enters linearly into the Hamiltonian (see eq. (\ref{HH})). Because the total mass of a liquid $\int dx \rho$ is conserved owing to the continuity equation, the term $w \rho$ contributes an additive constant to the Hamiltonian. In what follows, we set $w=0$. A simple corollary of (\ref{pot1})
\be\label{corr}
\int dx \left( \ell x_i \frac{\partial p}{\partial x_i} +V\right)=0
\ee
will prove useful below.

Infinitesimal form of the $\ell$--conformal Galilei transformations acting upon the fields $\rho(t,x)$, and $\upsilon_i (t,x)$ is presented in Appendix. Given a particular solution $(\rho(t,x),\upsilon_i (t,x))$ to the equations of motion (\ref{fin}), the pair $(\rho'(t,x),\upsilon'_i (t,x))$ displayed in Appendix provides another particular solution.

\vspace{0.5cm}

\noindent
{\bf 4. Conserved charges}\\

\noindent
As the next step, let us discuss conserved charges associated with the $\ell$--conformal Galilei symmetry of eqs. (\ref{fin}). Note that for arbitrary value of $\ell$, eqs. (\ref{fin}) involve a higher order material derivative. In this sense, they bear resemblance to higher derivative mechanics models studied in \cite{LSZ1,DH1,GK,AGM,AGKM,GM4,M}. As is known, for integer $\ell$ and $d>2$ the construction of an action functional for a higher--derivative mechanics appears problematic as the kinetic term turns into a total derivative \cite{GK,M}. Below we construct $SL(2,R)$--charges for a half--integer $\ell$ only, while the vector generators escape such a restriction. To a large extent, the analysis in this section is inspired by the instance of $\ell=\frac 12$ studied in \cite{JNPP}.

Let us start with the tower of vector generators. Taking into account the continuity equation and assuming $\upsilon_i$ to vanish on the boundary of a volume element, one can establish the relations
\be\label{supp3}
\frac{\partial}{\partial t} \left( \int d x \rho  \mathcal{D}^n \upsilon_i \right) =\int d x \rho \mathcal{D}^{n+1} \upsilon_i, \qquad
\frac{\partial}{\partial t} \left( \int d x \rho  x_i \right) =\int d x \rho \upsilon_i,
\ee
where $n$ is a natural number.
The generalised Euler equation $\rho  \mathcal{D}^{2\ell} \upsilon_i=-\frac{\partial p}{\partial x_i}$ then implies conservation of the momentum
\be\label{C0}
C^{(0)}_i=\int d x \rho \mathcal{D}^{2\ell-1} \upsilon_i,
\ee
provided the pressure $p$ (and hence the fluid density $\rho$) tends to zero at the boundary. Likewise, setting $n=2\ell-2$ in (\ref{supp3}), one gets
\be
\frac{\partial}{\partial t} \left( \int d x \rho  \mathcal{D}^{2\ell-2} \upsilon_i \right)=C^{(0)}_i.
\ee
Integrating both sides over the temporal variable $t$ and denoting a constant of integration by $-C^{(1)}_i$, one obtains a conserved charge associated with the Galilei boosts
\be
C^{(1)}_i=t C^{(0)}_i -\int d x \rho  \mathcal{D}^{2\ell-2} \upsilon_i.
\ee
Other vector generators are built by iterations\footnote{The structure relation $[H,C^{(n)}_i]=n C^{(n-1)}_i$ in (\ref{algebra}) is reproduces after the rescaling $k! C^{(k)}_i \to C^{(k)}_i.$}
\bea
&&
C^{(2)}_i=\frac{t^2}{2!} C^{(0)}_i-t C^{(1)}_i-\int d x \rho  \mathcal{D}^{2\ell-3} \upsilon_i,
\nonumber\\[2pt]
&&
C^{(3)}_i=\frac{t^3}{3!} C^{(0)}_i-\frac{t^2}{2!} C^{(1)}_i-t C^{(2)}_i-\int d x \rho  \mathcal{D}^{2\ell-4} \upsilon_i,
\nonumber\\[2pt]
&&
\dots
\nonumber\\[2pt]
&&
C^{(2\ell)}_i=\frac{t^{2\ell}}{(2 \ell) !} C^{(0)}_i-\frac{t^{2\ell-1}}{(2\ell -1)!} C^{(1)}_i-\dots-t C^{(2\ell-1)}_i-\int d x \rho  x_i,
\eea
where we used the notation $\mathcal{D}^{-1} \upsilon_i=x_i$.

Proceeding to conserved charges associated with the $SL(2,R)$--transformation, let us first attend to the energy--momentum tensor. Introducing the potential $V(\rho)$ specified by eq. (\ref{pot}) above, contracting the generalised Euler equation $\rho  \mathcal{D}^{2\ell} \upsilon_i=-\frac{\partial p}{\partial x_i}$ with $\upsilon_i$, and assuming $\ell$ to be half--integer\footnote{$n$ entering eq. (\ref{def}) and the equations below should not be confused with $n$ labeling the acceleration transformations in Sect. 2.}
\be\label{def}
\ell=\frac{2n+1}{2},
\ee
$n$ being a natural number,
one can bring the resulting expression to the form
\be\label{T00}
\partial_0 T^{00}+\partial_i T^{i0}=0,
\ee
which involves (no sum over repeated index  $n$)
\bea\label{emt}
&&
T^{00}=\rho \left(\frac{{(-1)^n}  \left( \mathcal{D}^n \upsilon_i \right) \left( \mathcal{D}^n \upsilon_i \right) }{2}+\sum_{k=0}^{n-1} {(-1)}^{k} \left( \mathcal{D}^k \upsilon_i \right) \left( \mathcal{D}^{2n-k} \upsilon_i \right)\right)+V(\rho),
\nonumber\\[2pt]
&&
T^{i0}=\rho \upsilon_i \left(\frac{{(-1)^n}  \left( \mathcal{D}^n \upsilon_i \right) \left( \mathcal{D}^n \upsilon_i \right)}{2}+\sum_{k=0}^{n-1} {(-1)}^{k} \left( \mathcal{D}^k \upsilon_i \right) \left( \mathcal{D}^{2n-k} \upsilon_i \right)+ V'(\rho) \right).
\eea
In obtaining (\ref{emt}), the identities
\be\label{supp4}
\rho \mathcal{D} A=\frac{\partial(\rho A)}{\partial t} +\frac{\partial (\rho \upsilon_i  A)}{\partial x_i}, \qquad \upsilon_i \frac{\partial p}{\partial x_i}=\frac{\partial V(\rho)}{\partial t}+\frac{\partial (\rho\upsilon_i V'(\rho))}{\partial x_i},
\ee
where $A(t,x)$ is an arbitrary function, which hold true due to the continuity equation, were used. Notice also a simple corollary of (\ref{supp4})
\be\label{supp5}
\frac{\partial}{\partial t} \left(\int dx \rho A \right) =\int dx \rho \mathcal{D} A,
\ee
which will prove useful below.

In a similar fashion, the generalised Euler equation $\rho \mathcal{D} \left( \mathcal{D}^{2\ell-1} \upsilon_i \right)+\frac{\partial p}{\partial x_i}=0$
can be used to build
\be\label{emt1}
T^{0i}=\rho  \mathcal{D}^{2\ell-1} \upsilon_i, \qquad T^{ji}=p \delta_{ji}+\rho \upsilon_j  \mathcal{D}^{2\ell-1} \upsilon_i,
\ee
which obey
\be\label{emt2}
\partial_0 T^{0i}+\partial_j T^{ji}=0.
\ee
When combined together, eqs. (\ref{emt}) and (\ref{emt1}) provide the energy--momentum tensor $T^{\mu\nu}$, $\mu=(0,i)$, $i=1,\dots,d$, associated with the equations of fluid dynamics (\ref{fin}) for a half--integer $\ell$. It is worth mentioning that $T^{\mu\nu}$ is not symmetric and eqs. (\ref{emt1}), (\ref{emt2}) hold for an arbitrary (i.e. not necessarily half--integer) value of $\ell$.

Now let us turn to conserved charges which link to the $SL(2,R)$ group. From (\ref{T00}), one finds the conserved energy (no sum over repeated index $n$)
\bea\label{HH}
&&
H=\int d x T^{00}
\nonumber\\[2pt]
&&
\quad
=\int d x \left(\rho \left(\frac{{(-1)^n}  \left( \mathcal{D}^n \upsilon_i \right) \left( \mathcal{D}^n \upsilon_i \right)}{2}+\sum_{k=0}^{n-1} {(-1)}^{k} \left( \mathcal{D}^k \upsilon_i \right) \left( \mathcal{D}^{2n-k} \upsilon_i \right) \right)+V(\rho)\right).
\eea

Rewriting $T^{00}$ in (\ref{emt}) as follows (no sum over repeated index $n$)
\bea
&&
T^{00}=\rho  \mathcal{D} \left(\frac 12 \sum_{k=0}^{n-1} {(-1)}^{k+n} (2k+1) \left(\mathcal{D}^{n-k-1} \upsilon_i \right) \left( \mathcal{D}^{n+k}\upsilon_i \right)+\ell x_i  \mathcal{D}^{2\ell-1} \upsilon_i \right)
\nonumber\\[2pt]
&&
\qquad \quad
+\ell x_i \frac{\partial p}{\partial x_i}+V(\rho),
\eea
where $\ell=\frac{2n+1}{2}$, and taking into account eqs. (\ref{corr}) and (\ref{supp5}), one gets a conserved charge associated with the dilatation (no sum over repeated index $n$)
\bea\label{DD}
&&
D=tH-\frac 12 \int dx \rho \left(  \sum_{k=0}^{n-1} {(-1)}^{k+n} (2k+1) \left(\mathcal{D}^{n-k-1} \upsilon_i \right) \left( \mathcal{D}^{n+k}\upsilon_i \right)+(2n+1) x_i  \mathcal{D}^{2n} \upsilon_i  \right).
\nonumber\\[2pt]
&&
\eea

Like $D$ was obtained by integrating both sides of $H=\int d x T^{00}$ over the temporal variable $t$, the conserved charge related to the special conformal transformation
\bea
&&
K=t^2 H-2t D-\int dx \rho \left(\sum_{k=0}^{n-1} {(-1)}^{k+n} (2k+1) \left( \frac{{(-1)^k}  \left( \mathcal{D}^{n-1} \upsilon_i \right) \left( \mathcal{D}^{n-1} \upsilon_i \right)}{2}
\right.
\right.
\nonumber\\[2pt]
&&
\qquad
\left.
+\sum_{s=0}^{k-1} {(-1)}^s \left(\mathcal{D}^{n-k+s-1} \upsilon_i \right) \left( \mathcal{D}^{n+k-s-1}\upsilon_i \right) \right)
\nonumber\\[2pt]
&&
\left.
\qquad
+(2n+1) \left(\frac{{(-1)}^n \left( \mathcal{D}^n x_i \right) \left( \mathcal{D}^n x_i \right)  }{2}+\sum_{s=0}^{n-1} {(-1)}^s \left( \mathcal{D}^s x_i \right) \left( \mathcal{D}^{2n-s} x_i \right) \right)
\right),
\eea
follows from (\ref{DD}) and the identities (no sum over repeated indices $n$ and $k$)
\bea
&&
 \left(\mathcal{D}^{n-k-1} \upsilon_i \right) \left( \mathcal{D}^{n+k}\upsilon_i \right)
\nonumber\\[2pt]
&& \quad
=\mathcal{D} \left(\frac{{(-1)^k}  \left( \mathcal{D}^{n-1} \upsilon_i \right) \left( \mathcal{D}^{n-1} \upsilon_i \right)}{2}+\sum_{s=0}^{k-1} {(-1)}^s \left(\mathcal{D}^{n-k+s-1} \upsilon_i \right) \left( \mathcal{D}^{n+k-s-1}\upsilon_i \right)  \right),
\eea
where $n\geq 1$ and $k\leq n-1$, and (no sum over repeated index $n$)
\be
x_i  \mathcal{D}^{2n+1} x_i=\mathcal{D} \left(\frac{{(-1)}^n \left( \mathcal{D}^n x_i \right) \left( \mathcal{D}^n x_i \right)  }{2}+\sum_{s=0}^{n-1} {(-1)}^s \left( \mathcal{D}^s x_i \right) \left( \mathcal{D}^{2n-s} x_i \right) \right).
\ee
In all the equations above, it is implied that a sum should be discarded whenever an upper bound is equal to $-1$.
In particular, choosing $\ell=\frac 12$ in the formulae above, one reproduces the result in \cite{JNPP}.

\vspace{0.5cm}

\noindent
{\bf 5. Conclusion}\\

\noindent
To summarise, in this work the equations of fluid dynamics were formulated, which hold invariant under the action of the $\ell$--conformal Galilei group. They included the conventional continuity equation, a higher order material derivative analogue of the Euler equation, and a suitable modification of the conventional equation of state. Conserved charges associated with the $\ell$--conformal Galilei symmetry transformations were built. It was demonstrated that, while the acceleration charges could be constructed both for integer and half--integer values of $\ell$, the $SL(2,R)$--charges and the energy--momentum tensor were straightforward to build for a half--integer $\ell$ only.

Turning to possible further developments, it would be interesting to construct a Hamiltonian formulation for the higher derivative fluid equations in Sect. 3. As was demonstrated in \cite{JNPP}, the $\ell=\frac 12$ Hamiltonain
\be
H=\int d x \left(\frac 12 \rho \upsilon_i \upsilon_i+V(\rho) \right),
\nonumber
\ee
gives rise to the continuity equation and the Euler equation, provided the Poisson bracket
\be
\{\rho(t,x),\upsilon_i (t,x') \}=-\frac{\partial \delta(x-x')}{\partial x_i}, \qquad \{\upsilon_i (t,x),\upsilon_j (t,x') \}=\frac{1}{\rho} \left(\frac{\partial\upsilon_j}{\partial x_i}-\frac{\partial\upsilon_i}{\partial x_j}  \right) \delta(x-x')
\nonumber
\ee
is introduced. A similar formalism for an arbitrary half--integer $\ell$ deserves a separate study. 

Explicit solutions to the generalised Euler equations and their dynamical stability are worth exploring as well. In particular, it would be interesting to understand whether the $2d$ variant of eqs. (\ref{fin}) admits generalised $n$--vortex solutions similar to those in \cite{AG}. 

Finally,
a possibility to use the equations in Sect. 3 within the context of the fluid/gravity correspondence (for a review see \cite{MR}) is worth exploring.

\vspace{0.5cm}

\noindent{\bf Acknowledgements}\\

\noindent
This work is supported by the Russian Foundation for Basic Research, grant No 20-52-12003.

\vspace{0.5cm}

\noindent
{\bf Appendix: Infinitesimal transformations}\\

\noindent
In this Appendix, we display infinitesimal form of the $\ell$--conformal Galilei transformations acting upon the coordinates $(t,x_i)$ and the fields $(\rho(t,x),\upsilon_i (t,x))$.

Substituting $\alpha=1$, $\delta=1$, $\gamma=0$  into eqs. (\ref{sl2}), (\ref{trr}), (\ref{trv}) and regarding $\beta$ as infinitesimal parameter, one obtains the infinitesimal form of the temporal translation
\begin{align}
&
t'=t+\beta, && x'_i=x_i,
\nonumber\\[2pt]
&
\rho'(t',x')=\rho(t,x), && \rho' (t,x)=\rho(t,x)-\frac{\partial \rho(t,x)}{\partial t} \beta,
\nonumber\\[2pt]
&
\upsilon'_i(t',x')=\upsilon_i (t,x), && \upsilon_i' (t,x)=\upsilon_i (t,x)-\frac{\partial \upsilon_i (t,x)}{\partial t} \beta.
\nonumber
\end{align}

Choosing $\alpha=e^{\frac{\lambda}{2}}$, $\delta=e^{-\frac{\lambda}{2}}$, $\beta=0$, $\gamma=0$, setting $\lambda$ to be infinitesimal parameter and Taylor expanding in $\lambda$ up to the first order, one gets the dilatation transformation
\begin{align}
&
t'=t+\lambda t, && x'_i=x_i+\lambda \ell  x_i,
\nonumber\\[2pt]
&
\rho'(t',x')=(1-\lambda \ell d) \rho(t,x), && \rho' (t,x)=(1-\lambda \ell d) \rho(t,x)-\lambda t \frac{\partial \rho(t,x)}{\partial t}-\lambda \ell x_i \frac{\partial \rho(t,x)}{\partial x_i},
\nonumber\\[2pt]
&
\upsilon'_i(t',x')=(1+\lambda (\ell-1))\upsilon_i (t,x), && \upsilon_i' (t,x)=(1+\lambda (\ell-1))\upsilon_i (t,x)-\lambda t \frac{\partial \upsilon_i(t,x)}{\partial t}
\nonumber\\[2pt]
&
&&
\qquad \qquad
-\lambda \ell x_j \frac{\partial \upsilon_i (t,x)}{\partial x_j}.
\nonumber
\end{align}

Infinitesimal form of the special conformal transformation is found by setting $\alpha=1$, $\delta=1$, $\beta=0$, $\gamma=-\sigma$, with infinitesimal $\sigma$, and Taylor expanding in $\sigma$ up to the first order
\begin{align}
&
t'=t+\sigma t^2, && x'_i=x_i+2\sigma \ell t  x_i,
\nonumber\\[2pt]
&
\rho'(t',x')=(1-2\sigma \ell d t) \rho(t,x), && \rho' (t,x)=(1-2\sigma \ell d t) \rho(t,x)-\sigma t^2 \frac{\partial \rho(t,x)}{\partial t}
\nonumber\\[2pt]
&
&& \qquad \qquad
-2\sigma \ell t x_i \frac{\partial \rho(t,x)}{\partial x_i},
\nonumber\\[2pt]
&
\upsilon'_i(t',x')=(1+2\sigma (\ell-1)t)\upsilon_i (t,x) && \upsilon_i' (t,x)=(1+2\sigma (\ell-1)t) \upsilon_i (t,x)-\sigma t^2 \frac{\partial \upsilon_i(t,x)}{\partial t}
\nonumber\\[2pt]
&
\qquad \qquad \quad +2 \sigma \ell x_i, && \qquad \qquad   -2\sigma \ell t x_j \frac{\partial \upsilon_i (t,x)}{\partial x_j}+2\sigma\ell x_i.
\nonumber
\end{align}

The acceleration transformations can be treated likewise and the result reads (no sum over repeated index  $n$)
\begin{align}
&
t'=t, && x'_i=x_i+a^{(n)}_i t^n,
\nonumber\\[2pt]
&
\rho'(t',x')=\rho(t,x), && \rho' (t,x)=\rho(t,x)-t^n a^{(n)}_i  \frac{\partial \rho(t,x)}{\partial x_i},
\nonumber\\[2pt]
&
\upsilon'_i(t',x')=\upsilon_i (t,x)+n a^{(n)}_i t^{n-1}, && \upsilon_i' (t,x)=\upsilon_i (t,x)-a^{(n)}_j t^n \frac{\partial \upsilon_i (t,x)}{\partial x_j}+n a^{(n)}_i t^{n-1},
\nonumber
\end{align}
where $a^{(n)}_i$, $n=0,\dots, 2\ell$, are infinitesimal vector parameters.

Introducing the variations
\be
t'=t+\delta t, \qquad x'_i=x_i+\delta x_i, \qquad  \rho'(t',x')=\rho(t,x)+\delta \rho(t,x), \qquad \upsilon'_i(t',x')=\upsilon_i (t,x)+\delta \upsilon_i (t,x),
\nonumber
\ee
where $\delta$ is regarded as the product of an infinitesimal parameter and a generator,
and computing the commutators $[\delta_1,\delta_2]$ acting upon both the coordinates and fields, one reproduces the algebra (\ref{algebra}).

\end{document}